\documentclass[preprint,12pt]{elsarticle}
\usepackage{amsmath}
\usepackage{epsfig}
\usepackage{amssymb}
\usepackage{dcolumn}
\usepackage{graphicx}
\usepackage{dcolumn}
\usepackage{color}
\usepackage{amssymb}

\journal{Optics Communications}

\begin{document}

\begin{frontmatter}

\title{Localized modes in two-dimensional Schr\"{o}dinger lattices with a
pair of nonlinear sites}

\author{Valeriy A. Brazhnyi}
\ead{brazhnyy@gmail.com}
\address{Centro de F\'{\i}sica do Porto, Faculdade de Ci\^encias, Universidade do
Porto, R. Campo Alegre 687, Porto 4169-007, Portugal}

\author{Boris A. Malomed}
\ead{malomed@post.tau.ac.il}
\address{Department of Physical Electronics, School of Electrical Engineering,
Faculty of Engineering, Tel Aviv University, Tel Aviv 69978, Israel}

\begin{abstract}
We address the existence and stability of localized modes in the
two-dimensional (2D) linear Schr\"{o}dinger lattice with two symmetric
nonlinear sites embedded into it, and a generalization for moderately
localized nonlinearity featuring two distinct symmetric maxima. The latter
setting admits a much greater variety of localized modes. Symmetric,
antisymmetric, and asymmetric discrete solitons are found, and a subcritical
bifurcation, accounting for the spontaneous symmetry breaking (SSB) of the
symmetric modes and transition to asymmetric ones, is identified. 
Existence and stability of more complex 2D solutions in the form of discrete symmetric and asymmetric vortices are also discussed.
\end{abstract}


\end{frontmatter}

\section{Introduction}

Discrete nonlinear systems, alias dynamical lattices, appear as basic models
in various fields of physics. Fundamental excitations supported by the
nonlinear lattice systems appear in the form of discrete solitons. In
particular, such solitons can be investigated in the framework of the
ubiquitous model based on the discrete nonlinear Schr\"{o}dinger (DNLS)
equation with the onsite cubic nonlinearity, or systems of such coupled
equations \cite{book}. Actually, the simplest variety of the DNLS equation
is the one with a pair of nonlinear sites embedded into the host linear
lattice \cite{Tsironis,we}, and similar models with a pair of nonlinear
sites side-coupled to a linear lattice \cite{Almas}.

It is well known that DNLS models of various types can be realized in
optics, in the form of transversely coupled arrays of parallel nonlinear
waveguides for optical \cite{big-review} and plasmonic \cite{plasmonics}
waves. On the other hand, the DNLS\ equation is a standard mean-field model
for Bose-Einstein condensates (BECs) trapped in deep periodic potentials
created, as optical lattices, by the interference of laser beams
illuminating the condensate in opposite directions, see original papers
which elaborated this setting for the atomic condensates \cite{BEC}, and
reviews \cite{BEC-review}. In particular, new techniques for engineering
various settings based on optical lattices have been developed recently \cite%
{Engineering-OL}. Further, the DNLS model with a few nonlinear sites
immersed into the linear host lattice can also be implemented in optics, by
embedding nonlinear cores into the linear waveguiding array, as discussed in
Ref.~\cite{Edwin}, and in BEC, by means of the Feshbach resonance \cite%
{Feshbach}, which can be induced locally by an external field in the
condensate trapped in the optical lattice. The latter possibility was
considered theoretically in various settings \cite{tiltedOLandFeshbach}, and
demonstrated experimentally in the ytterbium condensate, using an
appropriate laser field \cite{Yb}.

One of fundamental dynamical effects featured by nonlinear lattices is
spontaneous symmetry breaking (SSB). A typical manifestation of the SSB is
the transition from symmetric to asymmetric discrete solitons in the system
of two parallel DNLS chains, with the transverse linear coupling applied
uniformly \cite{Herring} or at single pair of sites \cite{Ljupco}. The
corresponding \textit{bifurcation}, i.e., the destabilization of the
symmetric solitons and emergence of asymmetric ones, is of the subcritical
and supercritical types in the former and latter cases, respectively (the
subcritical bifurcation, alias the phase transition of the first kind,
generates pairs of stable and unstable asymmetric states prior to the
destabilization of the parent symmetric one at the respective critical point
\cite{bif}; the supercritical bifurcation is a manifestation of the phase
transition of the second kind). Antisymmetric discrete solitons also exist
in these systems, but they do not undergo the bifurcation.

Another setting, which makes it possible to study the SSB of discrete
solitons in a single chain, is offered by the above-mentioned systems, based
on the linear chain with two nonlinear sites embedded into \cite{we} or
side-coupled \cite{Almas} onto it. In such models, the results can be
obtained in an analytical form. Also recently considered was the linear
chain with a pair of embedded sites with the quadratic
(second-harmonic-generating), rather than cubic, nonlinearity \cite{we2},
where analytical results can be obtained too.

In addition to the discrete systems, it is also possible to study solitons
and their SSB phenomenology in semi-discrete settings, with a pair of
localized elements, carrying the cubic \cite{Thawatchai} or quadratic \cite%
{Ady} nonlinearity, which are embedded into a continuous linear medium. The
latter setting may also be realized in both optics and BEC. A
two-dimensional (2D) version of this setting, with the cubic nonlinearity
concentrated at two separate spots, was recently considered too \cite%
{JModOpt}. Generally speaking, discrete and continuous settings with a few
localized nonlinear spots, or periodic arrays of such spots, belong to the
vast realm of \textit{nonlinear lattices} \cite{Barcelona}.

The subject of the present work is the study of 2D discrete solitons
supported by the nonlinearity concentrated at or around two sites embedded
into a host linear lattice, and the SSB phenomenology for such solitons. The
basic model where this setting can be implemented is the 2D DNLS network
with a pair of built-in nonlinear sites, or, in a more general form, with
localized nonlinearity that has maxima at two distinct sites. This setting
may be considered as a 2D version of the system introduced in Ref.~\cite{we}%
. It can be realized straightforwardly in BEC, using a deep optical lattice
and the locally concentrated action of the Feshbach resonance, as mentioned
above. In optics, the same setting may be built in the form of 2D
waveguiding arrays written as a permanent structure in bulk materials \cite%
{Jena}, with the strong nonlinearity in selected cores created by means of
dopants \cite{doping}. Alternatively, the effective nonlinearity of
particular cores in the waveguiding array can be enhanced simply be making
the effective cross-section area of this core smaller.

The model is formulated in Section II. Unlike its 1D counterpart the 2D
model, even in the case of the single nonlinear core, does not admit
analytical solutions. However, it gives rise to a much richer variety of
localized modes and transformations between them than was previously found
in the 1D geometry \cite{we}. Results of the numerical analysis are reported
in Sections III and IV for the cases of the strong and moderate localization
of the nonlinearity, respectively. In particular, the SSB occurs, in
contrast to the 1D system, for very broad solitons, and the character of the
bifurcation is always strongly subcritical. The paper is concluded by
Section IV.

\section{The model}

We consider the 2D DNLS equation as the model for the light transmission in
the square-shaped array of coupled optical waveguides, with propagation
distance $z$:
\begin{equation}
i\frac{du_{m,n}}{dz}+\Delta
_{m,n}u_{m,n}+W_{m_{0},n_{0}}(l)|u_{m,n}|^{2}u_{m,n}=0,  \label{1}
\end{equation}%
where the discrete 2D Laplacian is $\Delta
_{m,n}u_{m,n}=u_{m+1,n}+u_{m-1,n}+u_{m,n+1}+u_{m,n-1}-4u_{m,n}$. Two
nonlinear sites, embedded into the host linear lattice along the horizontal
direction, correspond to the following form of the local nonlinearity
strength in Eq. (\ref{1}):
\begin{equation}
W_{m_{0},n_{0}}(l)=\delta _{m_{0},n_{0}}+\delta _{m_{0}-l,n_{0}},
\label{nl_def}
\end{equation}%
where $(m_{0},n_{0})$ and $l$ are integer which define the coordinates of
the nonlinear cites and the distance between them.

It is also possible to consider two nonlinear sites embedded along the
diagonal direction, i.e., $W_{m_{0},n_{0}}(l)=\delta _{m_{0},n_{0}}+\delta
_{m_{0}-l,n_{0}-l}$. Below, we concentrate on the horizontal pair, as the
diagonal one yields similar results, but with a weaker interaction between
the nonlinear sites.

Stationary solutions to Eq.~(\ref{1}) with propagation constant $\omega $
are sought for as $u_{m,n}(z)=e^{i\omega z}U_{m,n}$, where stationary field $%
U_{m,n}$ (which is real for fundamental solitons, and complex for vortices)
obeys equation%
\begin{equation}
\omega U_{m,n}=\Delta _{m,n}U_{m,n}+W_{m_{0},n_{0}}(l)|U_{n}|^{2}U_{n}.
\label{U}
\end{equation}%
Stationary solutions are characterized by their norm (i.e., total power in
the case of the array of optical waveguides),
\begin{equation}
N=\sum_{m,n=-\infty }^{+\infty }|U_{m,n}|^{2}.  \label{N}
\end{equation}

\begin{figure}[h]
\epsfig{file=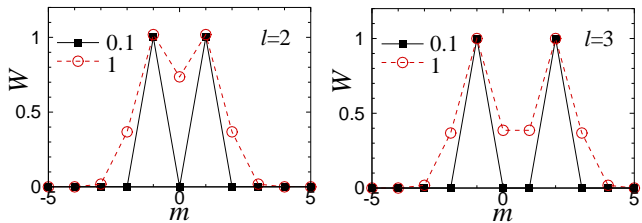,width=9cm}
\caption{(Color online) The cross section of the nonlinearity-modulation
profile, $W_{m,n=0}$, for the cases of strong ($\Delta W=0.1$) and moderate (%
$\Delta W=1$) localization [see Eq.~(\protect\ref{eq:delta})], and two
values of the integer distance between the nonlinear sites, $l=2$ and $3$.}
\label{W_01_1}
\end{figure}

As a generalization of the model with the two-site structure (\ref{nl_def}),
we will also consider the one with the Gaussian profile of the localized
double-spot nonlinearity,
\begin{equation}
\delta _{m_{0},n_{0}}\rightarrow \exp \left( -\frac{%
(m-m_{0})^{2}+(n-n_{0})^{2}}{(\Delta W)^{2}}\right) ,  \label{eq:delta}
\end{equation}%
in which the strong and moderate localization correspond, respectively, to $%
\Delta W\ll 1$ and $\Delta W\gtrsim 1$, as shown in Fig.~\ref{W_01_1}. In
the experiment, the Gaussian nonlinearity-modulation profile (\ref{eq:delta}%
) may be realized by means of the respective distribution of the density of
the resonant nonlinearity- inducing dopant, or by fabricating the array with
the corresponding distribution of the cores' effective cross-section area.

\section{Modes supported by the strong localization of the nonlinearity ($%
\Delta W=0.1$)}

\begin{figure}[h]
\epsfig{file=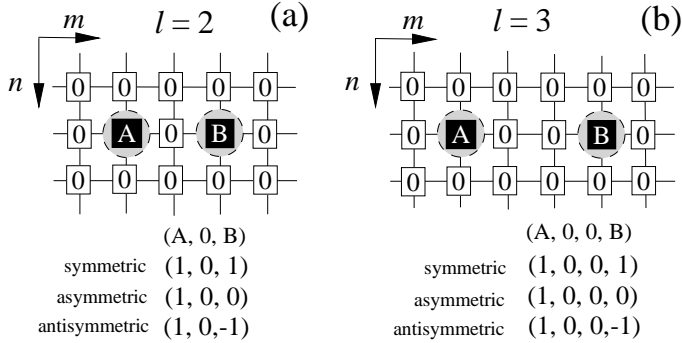,width=9cm}
\caption{Top row is schematic distribution of exited sites in the 2D
lattice. Gray circles show schematic localization of nonlinear spots. A and
B stand for amplitudes at the nonlinear spots. In the bottom the modes of
symmetric, asymmetric, and antisymmetric types are schematically shown in
the anticontinuum limit; they are localized on two nonlinear sites separated
by distance $l=2$ or $3$.}
\label{first}
\end{figure}

Unlike its 1D counterpart, the present model, even in its simplest form,
with a single or two nonlinear sites embedded into the linear host medium
[see Eq.~(\ref{nl_def})], does not admit an analytical solution for
stationary modes. Therefore, the starting point of the analysis is the
anticontinuum limit (ACL) \cite{book}, which corresponds, in the present
notation, to $\omega \rightarrow \infty $. In this limit, all the modes are
built as configurations with nonzero amplitudes set only at sites carrying
the strong nonlinearity. Such modes of the symmetric, asymmetric and
antisymmetric types are schematically shown in Fig.~\ref{first}, for $l=2$
and $l=3$.

\begin{figure}[h]
\epsfig{file=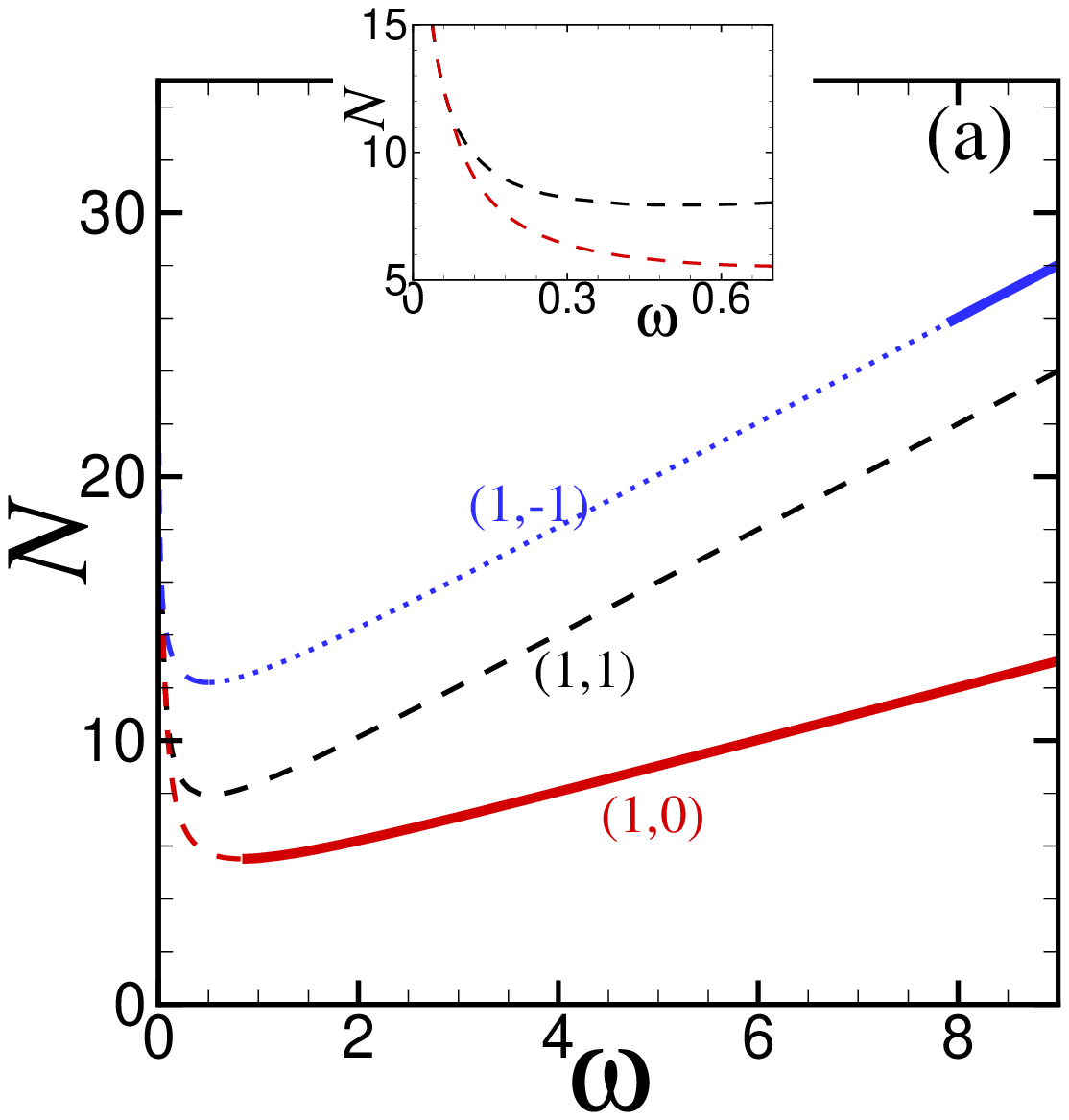,width=5cm}%
\epsfig{file=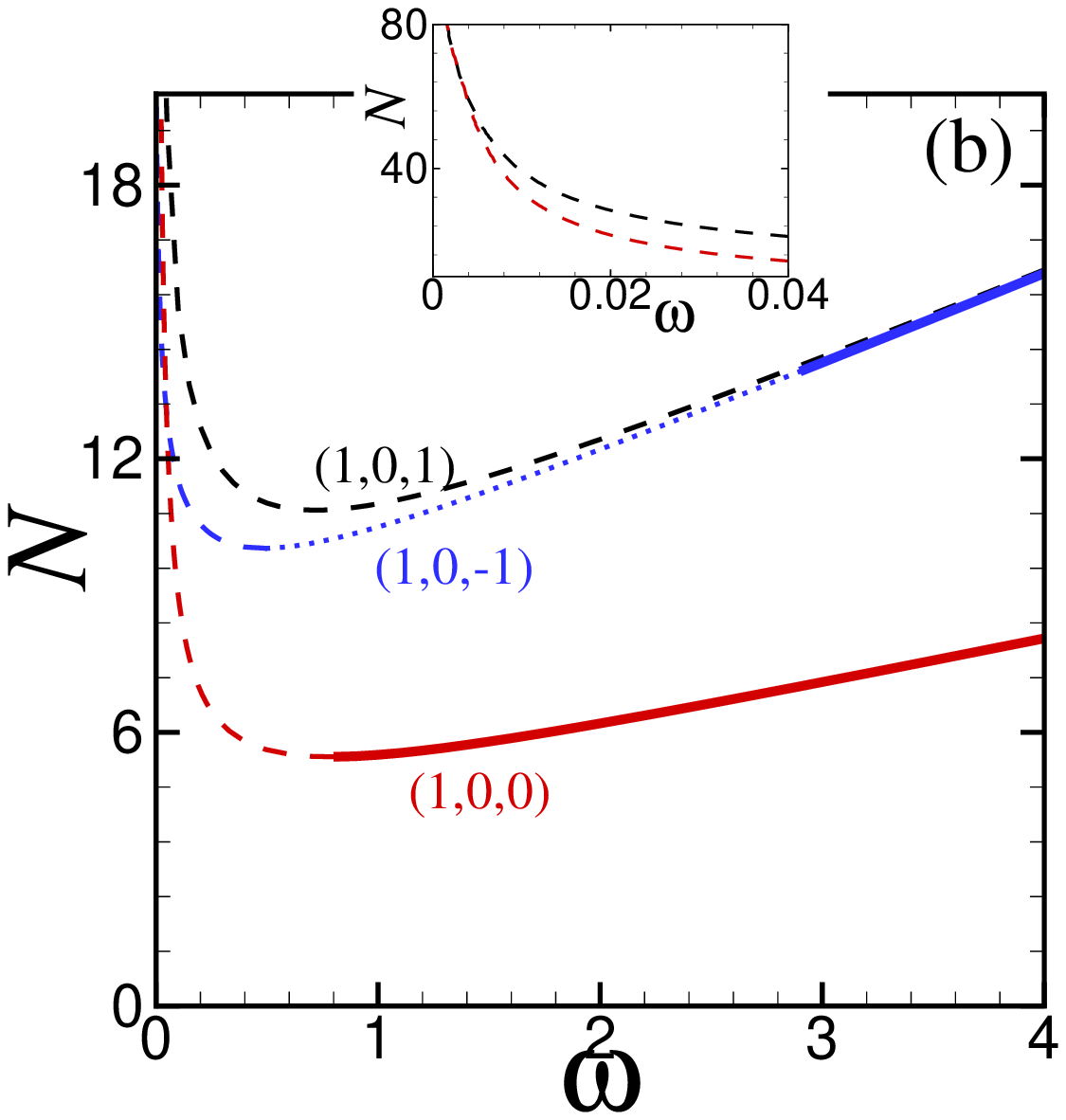,width=5cm}%
\epsfig{file=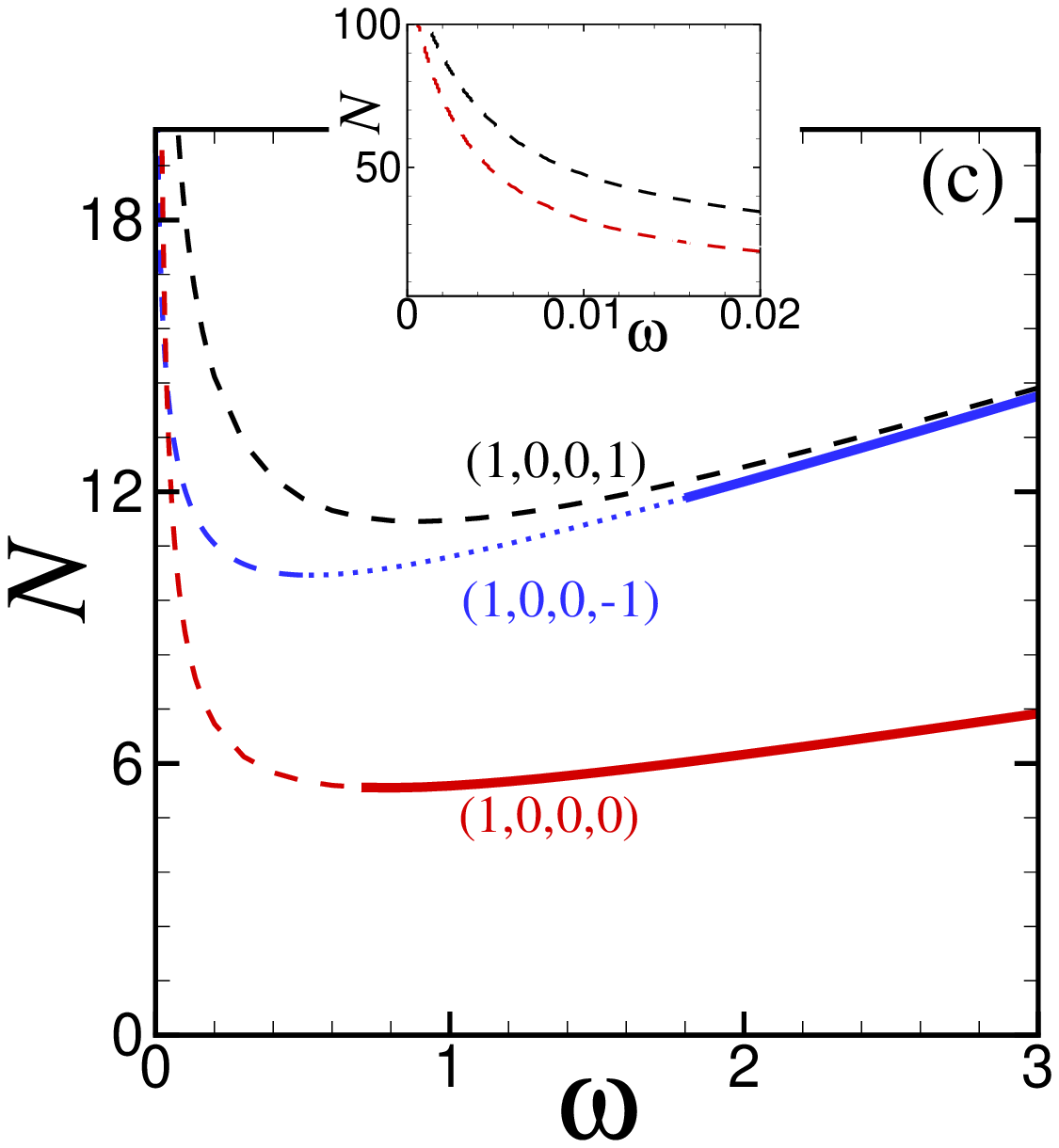,width=5cm}
\caption{(Color online) Curves $N(\protect\omega )$, the total power (norm)
versus the propagation constant [see Eq.~(\protect\ref{N})], for families of
symmetric (black), asymmetric (red) and antisymmetric (blue) modes. (a), (b)
and (c) correspond to distances $l=1,2$ and $3$ between the strongly
localized nonlinear sites, with $\Delta W=0.1$. Here and in similar figures
following below, solid, dotted and dashed lines represent stable,
oscillatory-unstable and strongly unstable branches of the solutions,
respectively. Insets zoom areas in which the SSB bifurcation occurs.}
\label{Nw_strong}
\end{figure}

In the following, various modes are designated according to the ACL seeds
from which they stem. The curves which represent the so found symmetric,
asymmetric and antisymmetric modes by means of dependencies $N(\omega )$ are
shown in Fig.~\ref{Nw_strong}. In contrast to the 1D model \cite{we} and 2D continuous NLS model with periodic potential \cite{Yang2008,Dohnal2009}, the norm of the localized modes diverges close to the phonon band, where the
modes undergo delocalization, i.e., may be considered in the continuum
approximation, with the nonlinear term multiplied by the $\delta $-function
of the spatial coordinates. The continual 1D Schr\"{o}dinger equation with
such a nonlinear term gives rise to a family of solitons with a finite norm
\cite{Thawatchai}, while such solutions do not exist in the 2D case, which
explains the above-mentioned divergence.

At $l=1$, i.e., in the 2D linear host lattice with the adjacent pair of
embedded nonlinear sites, the SSB, which gives rise to asymmetric localized
mode ($1,0$) from the symmetric one ($1,1$), occurs in strongly delocalized
states with large norms [see Fig.~\ref{Nw_strong}(a)]. The respective
bifurcation is subcritical, hence the emerging branch of the asymmetric
discrete solitons is unstable. With the decrease of $N$, it attains a
turning point, corresponding to a minimum of $N$, at which the branch
becomes stable. The stabilization of the asymmetric branch at the turning
point, where it changes the sign of the slope, from $dN/d\omega <0$ to $%
dN/d\omega >0$, exactly agrees with the Vakhitov-Kolokolov (VK) criterion
\cite{VK}.

Increasing distance $l$ between the embedded nonlinear sites from $1$ to $3$%
, we observe that the SSB takes place for the modes which are so broad that
they cannot be fitted to the computation domain. For the sake of the
comparison with the 1D model~\cite{we}, it is relevant to mention that the
latter one also gives rise to the subcritical SSB bifurcation for $l\geq 2$,
while, just for $l=1$, the respective bifurcation is supercritical in the 1D
system~\cite{we2}, on the contrary to the results reported here. The
difference in the character of the bifurcation (super/sub-critical) between
the 1D and 2D settings with $l=1$ may be explained by the general trend of
the SSB bifurcations to switch to the subcritical type with the increase of
the dimension (for instance, the bifurcation driven by the cubic term is
always supercritical in the zero-dimensional model, but tends to become
subcritical in 1D). For the model with exactly two nonlinear sites, i.e., $%
\Delta W=0$, which corresponds to Eq.~(\ref{nl_def}), the situation is
virtually the same as displayed here for $\Delta W=0.1$.

The stability of the modes along solution branches has been checked through
the computation of stability eigenvalues for small perturbations. The
perturbed solution was looked as
\begin{equation}
u_{m,n}(z)=\left[ U_{m,n}^{(s)}+a_{m,n}(z)\right] e^{i\omega z},
\label{pert}
\end{equation}%
where $a_{m,n}(z)=p_{m,n}e^{i\lambda z}+q_{m,n}^{\ast }e^{-i\lambda ^{\ast
}z}$ is an eigenmode of small perturbations with (generally, complex)
eigenvalue $\lambda $, added to stationary solution $U_{m,n}^{(s)}$ (the
asterisk stands for the complex conjugate). Substituting expression (\ref%
{pert}) into Eq. (\ref{1}) leads to the linearized eigenvalue problem, that
can be written in the following matrix form,
\begin{equation}
\left(
\begin{array}{cc}
\hat{L}_{m,n}-\lambda  & \hat{L}_{m,n}^{\prime } \\
-(\hat{L}_{m,n}^{\prime })^{\ast } & -\hat{L}_{m,n}-\lambda
\end{array}%
\right) \left(
\begin{array}{c}
p \\
q%
\end{array}%
\right) =0  \label{matrix}
\end{equation}%
where operators are $\hat{L}_{m,n}=-\omega +\Delta
_{m,n}+2W_{m,n}|U_{m,n}^{(s)}|^{2}$ and $\hat{L}_{m,n}^{\prime
}=W_{m,n}\left( U_{m,n}^{(s)}\right) ^{2}$. The stability of the solutions
was then analyzed through the calculation of eigenvalues of Eq. (\ref{matrix}%
), and verified by direct simulations of the perturbed evolution.

In this way, it was found that unstable modes designated in Fig.~\ref%
{Nw_strong} may feature different types of the instability. In particular,
as concerns the symmetric and antisymmetric modes shown in Fig.~\ref%
{Nw_strong}, the analysis demonstrates that the symmetric mode is strongly
unstable due to the presence of purely imaginary eigenvalues, irrespective
of the compliance with the VK criterion (recall that it gives only a
necessary but not sufficient stability condition). On the other hand, the
antisymmetric mode of the ($1,-1$) type, which is unstable at small $\omega $
to oscillatory perturbations, corresponding to a pair of complex
eigenvalues, transforms into an oscillatory unstable state after the turning
point. Then, it becomes stable at some critical propagation constant, whose
value decreases with the increasing of distance $l$ between the nonlinear
sites. This difference is also visible in the dynamical evolution of the
corresponding modes, with the strong instability of mode ($1,1$) developing
much faster than in the oscillatory instability of mode ($1,-1$) (not shown
here in detail).

\begin{figure}[h]
\epsfig{file=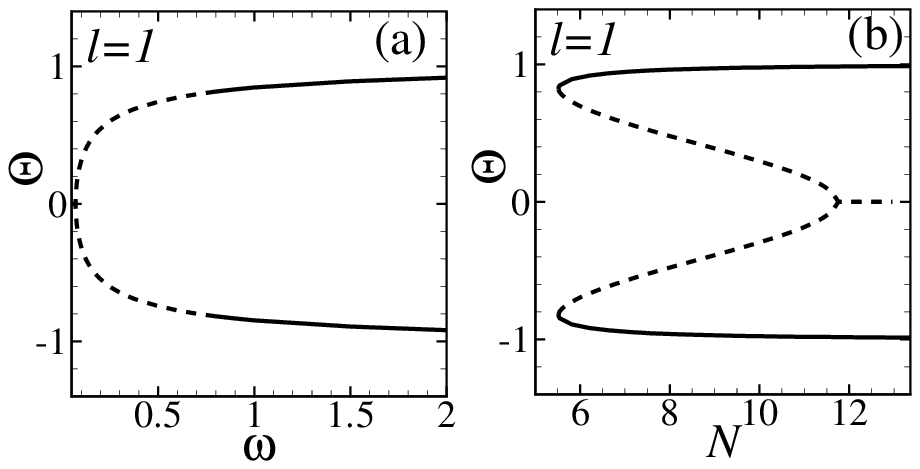,width=5cm}%
\epsfig{file=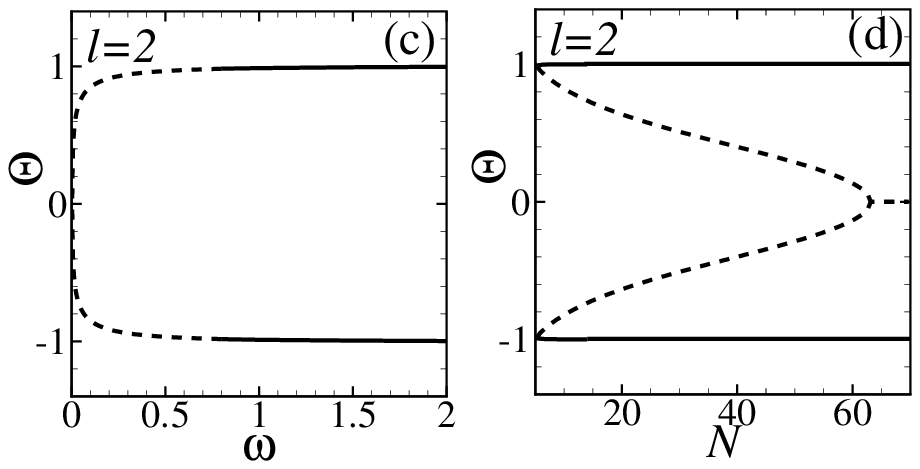,width=5cm}%
\epsfig{file=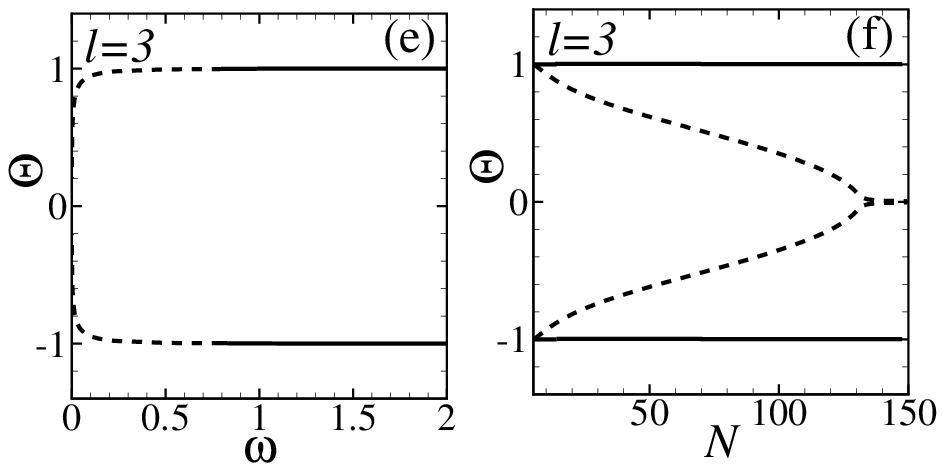,width=5cm}
\caption{Asymmetry parameter (\protect\ref{Theta}) of the mode (1,0), as a
function of $\protect\omega $ in (a), (c) and (e), and as a function of $N$
in (b), (d) and (f), for different distances between the nonlinear sites
(indicated in each plot) in the case of the strong localization of the
nonlinearity, $\Delta W=0.1$, for the same solutions which are shown in Fig.~%
\protect\ref{Nw_strong}. }
\label{Theta_Nw_strong}
\end{figure}

The SSB is characterized by the asymmetry parameter, which is defined as
\begin{equation}
\Theta \equiv \frac{U_{m_{0},n_{0}}^{2}-U_{m_{0}-l,n_{0}}^{2}}{%
U_{m_{0},n_{0}}^{2}+U_{m_{0}-l,n_{0}}^{2}},  \label{Theta}
\end{equation}
and may be displayed as a function of $\omega $ or $N$. The respective
results for strongly localized nonlinearity and for different distances
between the nonlinear sites are presented in Fig.~\ref{Theta_Nw_strong}. It
is seen that the SSB is \textit{strongly subcritical}, with the unstable
branches of the asymmetric modes going a long way in the direction of
decreasing $N$, and only afterwards the branches turn back, recovering the
stability of \emph{extremely asymmetric modes}, with values of $\Theta $
very close to $1$.

\begin{figure}[h]
\epsfig{file=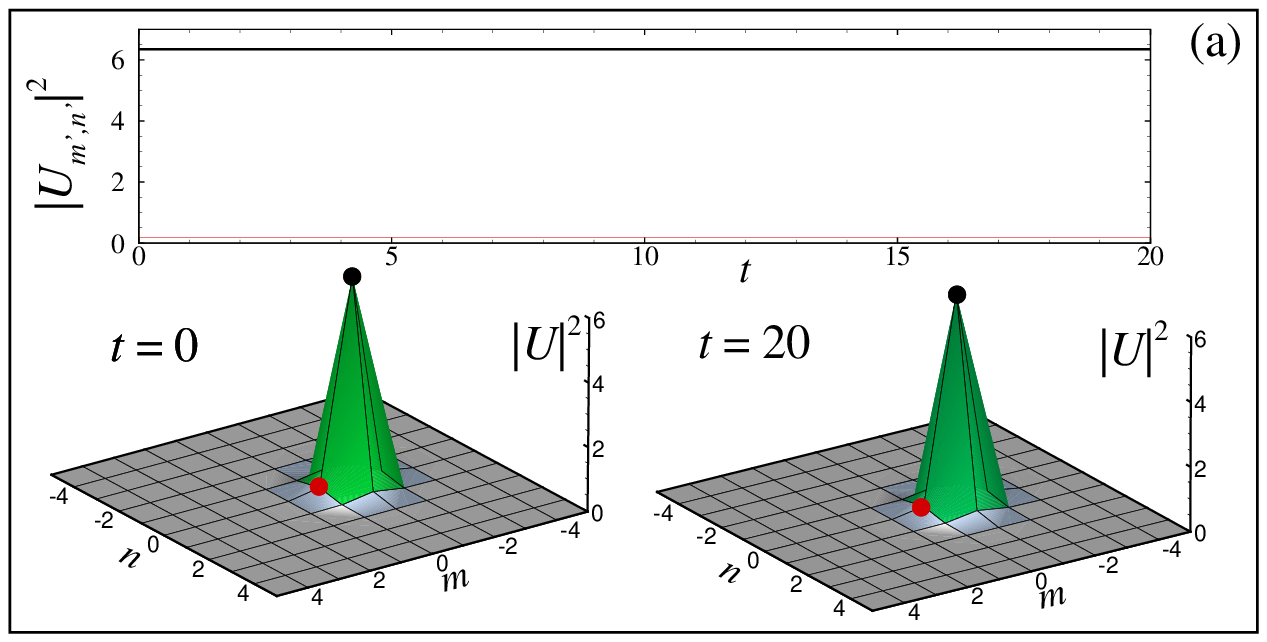,width=7cm}\hspace{1cm}%
\epsfig{file=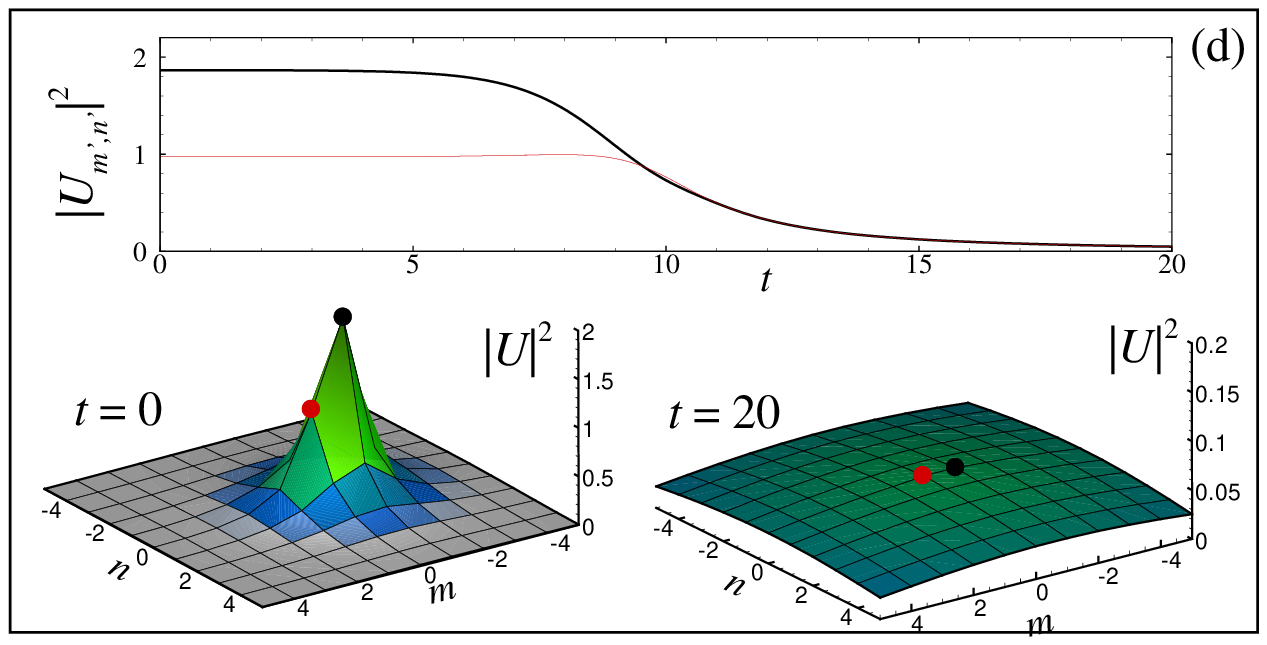,width=7cm} %
\epsfig{file=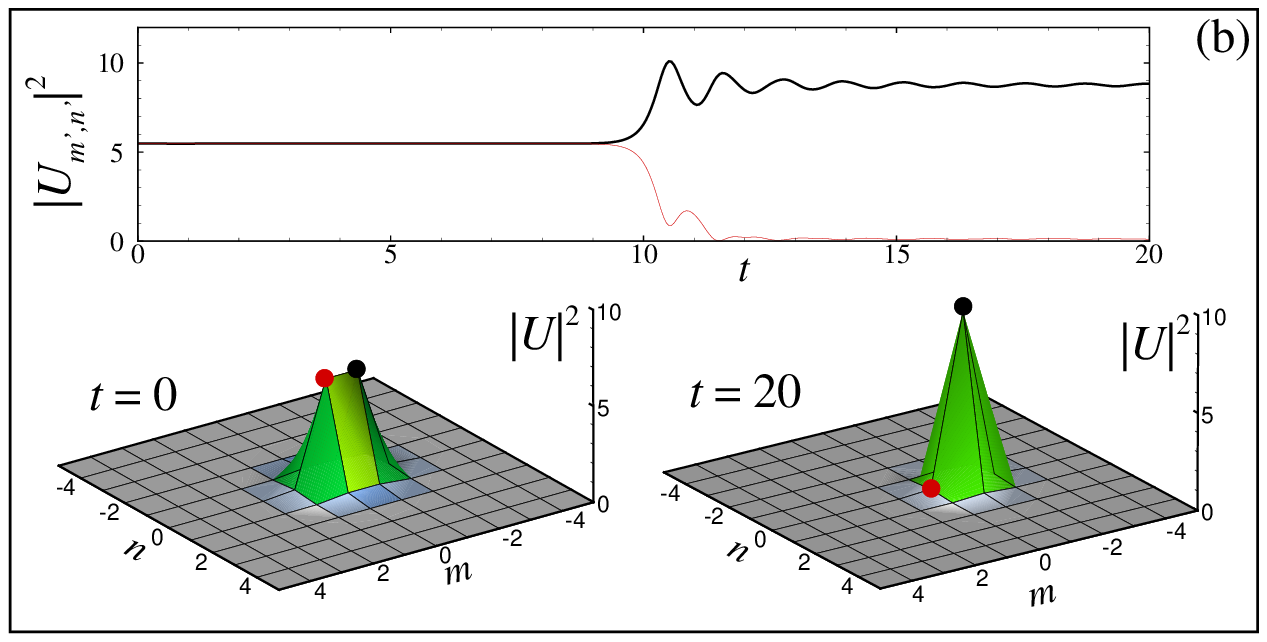,width=7cm}\hspace{1cm}%
\epsfig{file=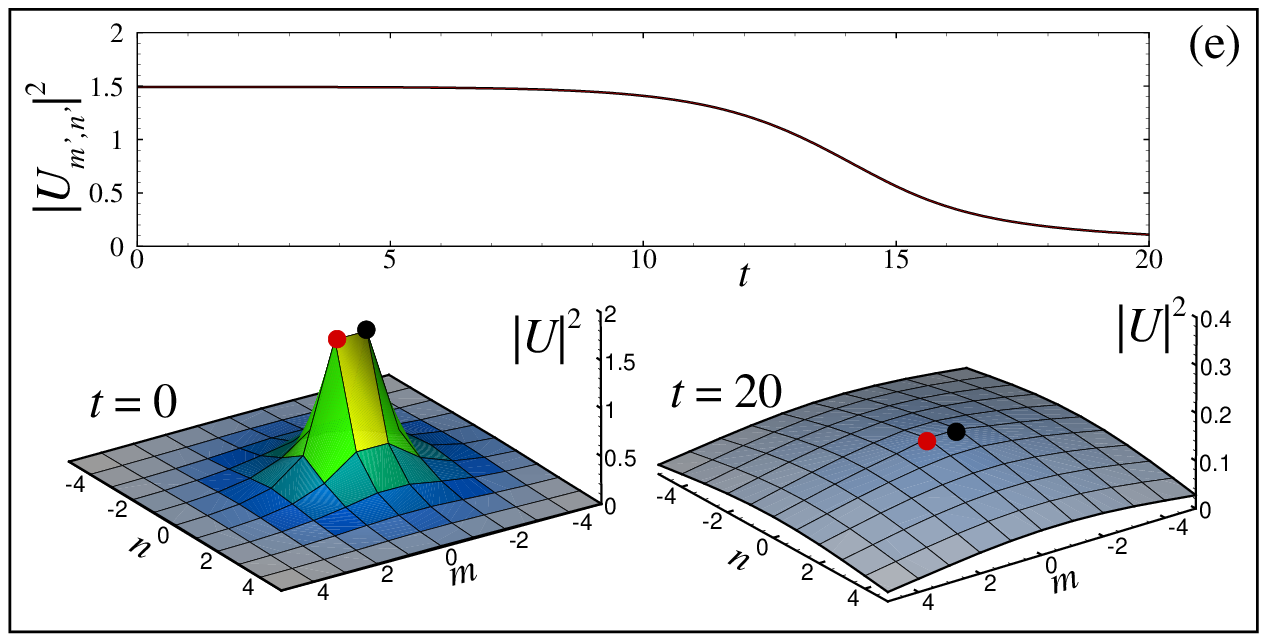,width=7cm} %
\epsfig{file=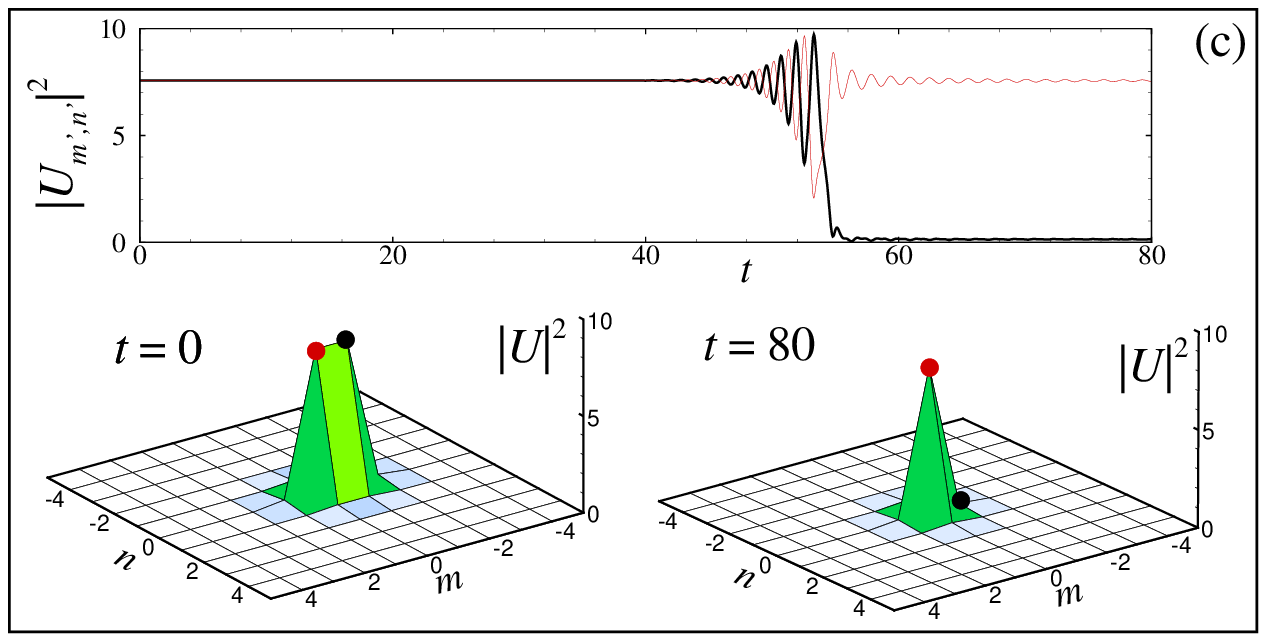,width=7cm}\hspace{1cm}%
\epsfig{file=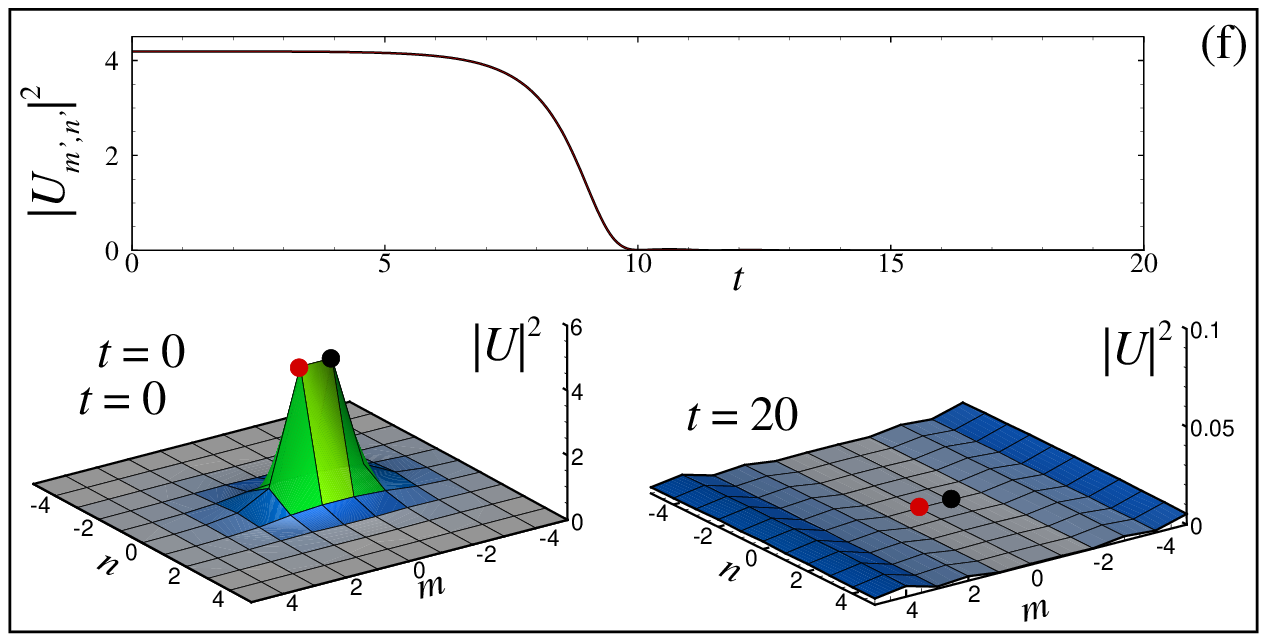,width=7cm}
\caption{(Color online) The evolution of asymmetric [(a), (d)], symmetric
[(b), (e)] and antisymmetric [(c), (f)] localized modes with $l=1$. In each
subplot the dynamics of two initially excited sites $|U_{m^{\prime
},n^{\prime }}|$, indicated by black and red dots on the initial ($t=0$) and
final ($t=20$) profiles, is shown. The left column corresponds to $\protect%
\omega =3$, and the right one to $\protect\omega =0.1$. The only stablemode
is the asymmetric one shown in (a) at $\protect\omega =3$, the other
examples being unstable.}
\label{dyn_prof}
\end{figure}

Dynamical properties of the localized modes are further illustrated by Fig.~%
\ref{dyn_prof}, which displays typical initial and final profiles, at $t=0$
and $t=20$, respectively, of symmetric, asymmetric and antisymmetric
localized solutions, as well as the evolution of the initially excited
sites. The figure also demonstrates that, at relatively large values of $%
\omega $, at which the asymmetric modes are strongly localized and stable,
their symmetric and antisymmetric counterparts are unstable against
symmetry- (or antisymmetry-) breaking perturbations, spontaneously
transforming into the asymmetric mode. On the other hand, at small $\omega $%
, which implies that the discrete solitons are broad, they are unstable
against full delocalization, as shown in the right-hand side of Fig.~\ref%
{dyn_prof}.

\section{Modes supported by the moderately localized nonlinearity ($\Delta
W=1$)}

\begin{figure}[h]
\epsfig{file=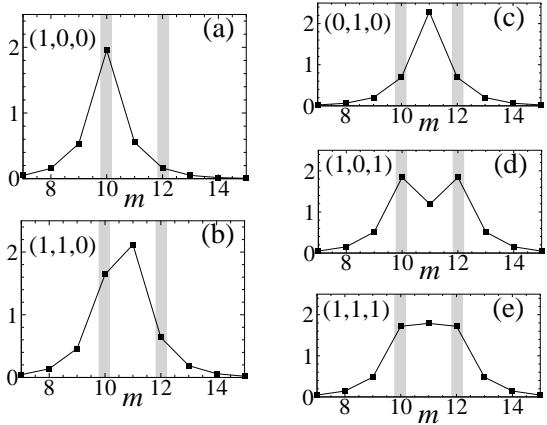,width=8cm}
\caption{Modes supported by the moderately localized nonlinearity ($\Delta
W=1$) at $l=2$ and $\protect\omega =1$. }
\label{modes_weak}
\end{figure}

The increase of the area of the nonlinearity localization, i.e., the
increase of $\Delta W$ in Eq.~(\ref{eq:delta}), makes it possible to build a
larger variety of 2D localized modes in the ACL, as shown in Figs.~\ref%
{first_weak} and \ref{modes_weak}. In this section we study modes for $%
\Delta W=1$. For example, in the case of $l=2$ the system supports a mode of
type ($0,1,0$), which does not exist in the limit of the strong
localization. This is possible because in the ACL the amplitude at each site
is related to the propagation constant and the magnitude of the nonlinear
coefficient as follows: $U_{n}=\{\pm \sqrt{\omega /W_{n}},0\}$. In the case
of the strong localization of $W_{n}$, one can construct only ACL modes
localized on each of the nonlinear sites. On the other hand, with moderate
localization of the nonlinearity, as in the case of $\Delta W=1$, it is
possible to excite modes on sites which are located off the center of the
nonlinear spot. For example, in the case of $l=2$ and $\Delta W=1$, the
nonlinearity at the midpoint between the two nonlinear sites is $\approx 0.75
$, thus allowing one to construct the ACL mode ($0,1,0$).

\begin{figure}[h]
\epsfig{file=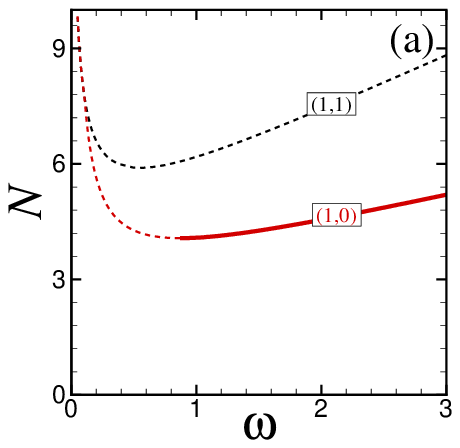,width=4cm}%
\epsfig{file=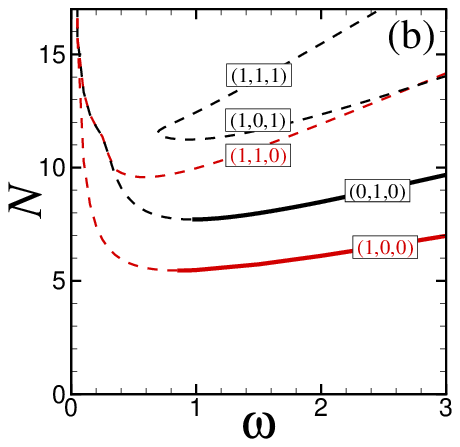,width=4cm}%
\epsfig{file=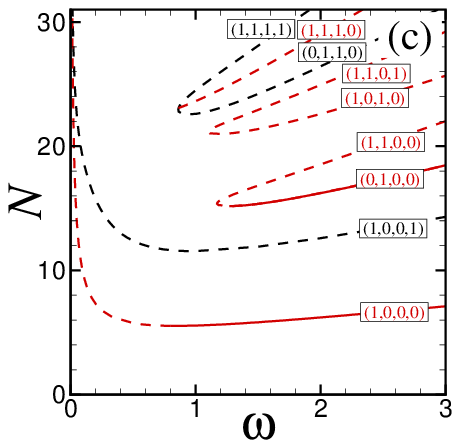,width=4cm} %
\epsfig{file=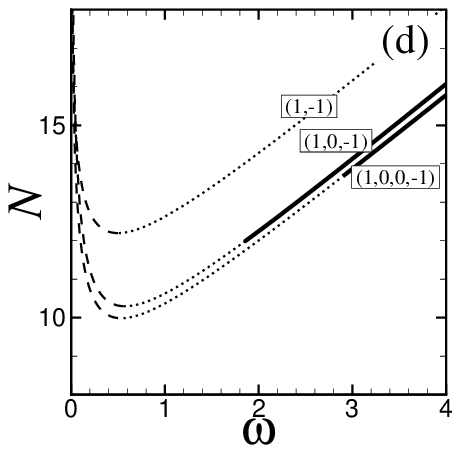,width=4cm}
\caption{(Color online) Curves $N(\protect\omega )$ for families of
localized symmetric, asymmetric and antisymmetric modes, supported by the
moderately localized nonlinearity, with $\Delta W=1$, for $l=1,\ 2$ and $3$.
In (a)-(c) the black and red colors correspond to symmetric and asymmetric
modes, respectively. Panel (d) shows the families of antisymmetric modes for
different distances $l$. }
\label{Nw_weak}
\end{figure}

The corresponding branches of symmetric, asymmetric and antisymmetric
localized modes are displayed in Fig.~\ref{Nw_weak}.

\begin{figure}[h]
\epsfig{file=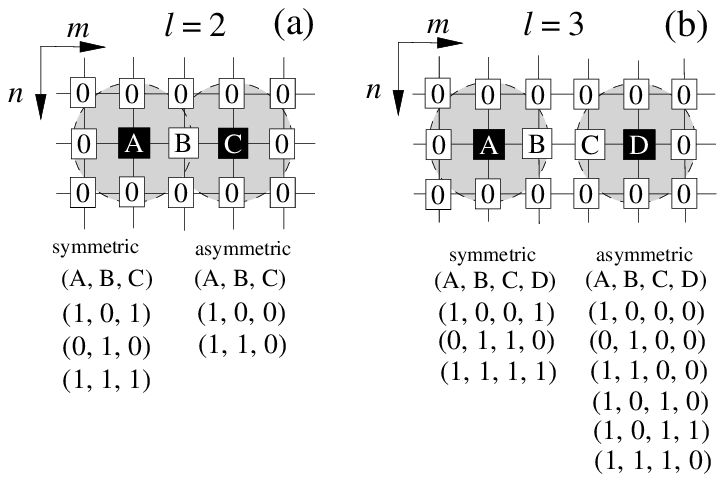,width=9cm}
\caption{Modes supported by the moderately localized nonlinearity in the
anticontinuum limit, for $l=2$ and $3$. Gray circles schematically indicate
the localization of the nonlinearity.}
\label{first_weak}
\end{figure}

This diagram is much richer than its counterpart in the case of the strong
localization, cf. Fig.~\ref{first}. In particular, two new branches of the
solutions [symmetric ($0,1,0$) and asymmetric ($1,1,0$)] can be built in the
ACL. In accordance with what said above, the partial delocalization of the
nonlinearity makes it possible to construct ACL solutions that are localized
not on the sites carrying the maximum nonlinearity, but on neighboring sites
where the nonlinearity is also tangible, as shown in Fig.~\ref{W_01_1}.

\begin{figure}[h]
\epsfig{file=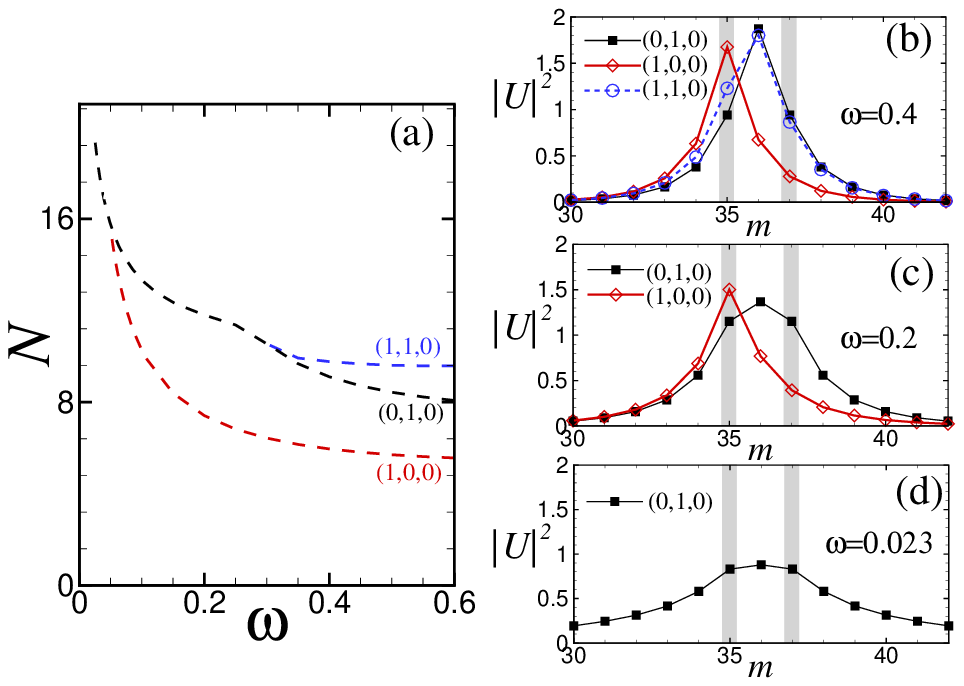,width=9cm}
\caption{(Color online) In (a) the existence curves $N(\protect\omega )$
from Fig.~\protect\ref{Nw_weak}(b) are shown in the interval of $\protect%
\omega \in \lbrack 0,0.6]$. In (b)-(d), the profiles of the solutions at
different propagation constants $\protect\omega =0.4;0.2$ and $0.023$ are
displayed. The gray stripes shows positions of the nonlinear spots.}
\label{bifurc_weak}
\end{figure}

In the case of $l=1$ and weak localization, the situation is very similar to
that corresponding to the strong localization. New types of bifurcations
(actually, a bifurcation cascade) are found at larger $l$. For example, in
the case of $l=2$, on the contrary to the situation corresponding to the
strong localization, symmetric mode ($1,0,1$) does not have a
small-amplitude limit, and disappears through a saddle-node bifurcation
involving another symmetric mode ($1,1,1$) (which does not exist in the
strong-localization model with $\Delta W=0.1$). The situation is different
too in the small-amplitude limit (for small $\omega $). In Fig. \ref%
{bifurc_weak}(a), the existence curves for the symmetric and asymmetric
solutions are presented in the region of small $\omega $. In panel (b) the
profiles of symmetric ($0,1,0$) and asymmetric ($1,0,0$) and ($1,1,0$) modes
are presented, before the first bifurcation. With the decrease of $\omega $,
the first pitchfork bifurcation occurs between asymmetric mode ($1,1,0$) and
symmetric one ($0,1,0$), transforming into a symmetric mode [see
corresponding profiles at $\omega =0.2$ in Fig.~\ref{bifurc_weak}(c)].
Later, symmetric mode ($0,1,0$) undergoes another pitchfork bifurcation with
asymmetric mode ($1,0,0$). The final profile of the symmetric mode is shown
in Fig.~\ref{bifurc_weak}(d).

By further increasing the distance between the nonlinear sites, $l$, the set
of solution branches becomes even more complex. For instance, the system
with $l=3$ shows a number of saddle-node bifurcations between modes ($1,1,0,1
$) and ($1,0,1,0$), ($1,1,0,0$) and ($0,1,0,0$). Also, a more complex
cascade of bifurcations occurs between modes ($1,1,1,1$) and ($1,1,1,0$)
(pitchfork) and, further, with mode ($0,1,1,0$) (saddle-node). The linear
stability analysis, displayed in Fig.~\ref{Nw_weak}, shows some noteworthy
effects, such as an additional stable asymmetric branch ($0,1,0,0$).

\section{Vortices supported by the localization of the nonlinearity}

Dealing with the 2D model, it is natural too to explore symmetric and
asymmetric vortices supported by the localized nonlinearity. In contrast to
the localized modes without vorticity, the localization strength plays a
crucial role for them \cite{vortex,book}.

\begin{figure}[h]
\epsfig{file=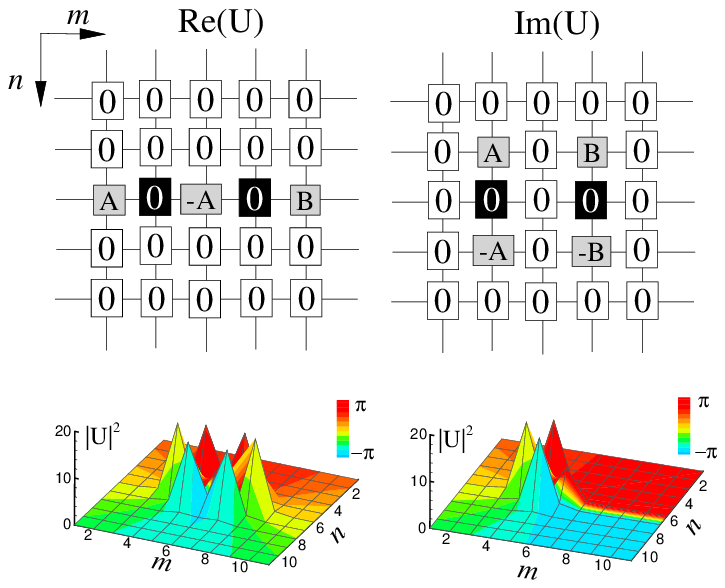,width=9cm}
\caption{(Color online) Top row shows a schematic distribution of the exited
sites in the real and imaginary parts of the vortex solution in the ACL. In
the bottom row, the profiles of $|U_{m,n}|^{2}$ are shown for the symmetric
(left, and asymmetric (right) vortices. The color gradient shows the phase
distribution. Parameters are $\protect\omega =2$, $l=2$, $\Delta W=1$.}
\label{vort_l2}
\end{figure}

\begin{figure}[h]
\epsfig{file=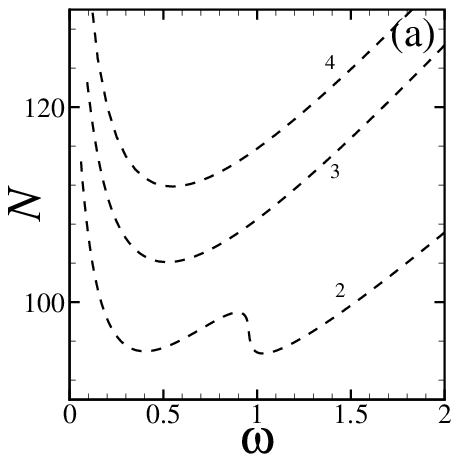,width=5cm}\epsfig{file=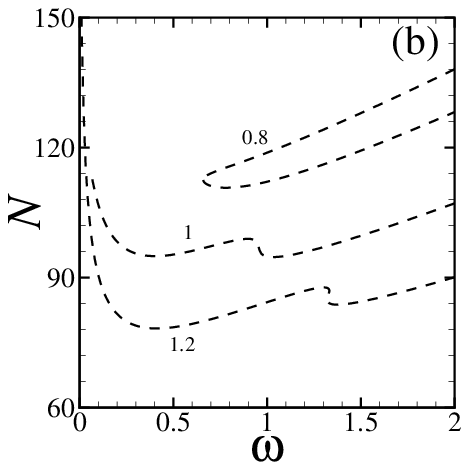,width=5cm}
\caption{Curves representing families of symmetric vortices with different
combination of the width of the nonlinear spots, $\Delta W$, and distance
between them, $l$. In (a) $\Delta W=1$ and $l=2,3,4$, while in panel (b) $l=2
$ and $\Delta W=0.8;1;1.2$.}
\label{vort_exist}
\end{figure}

We are looking for vortical solution to the Eq.~(\ref{U}) starting from the
ACL, as it was done in the previous section. The initial distribution of the
exited sites in the ACL, designated as $\pm A,\pm B$, and the final shape of
the symmetric (with $A=B=1$) and asymmetric ($A=1$ and $B\ll 1$) vortices,
represented by $|U_{m,n}|^{2}$ with $\Delta W=1$ and distance $l=2$, are
shown in Fig.~\ref{vort_l2}. In Figs.~\ref{vort_exist} and \ref%
{vort_exist_as}, the existence of the symmetric and asymmetric vortices for
different distances between nonlinear sites $l$ with fixed $\Delta W=1$
[panel (a)], and different width $\Delta W$ with fixed $l=2$ [panel (b)],
are presented.

\begin{figure}[h]
\epsfig{file=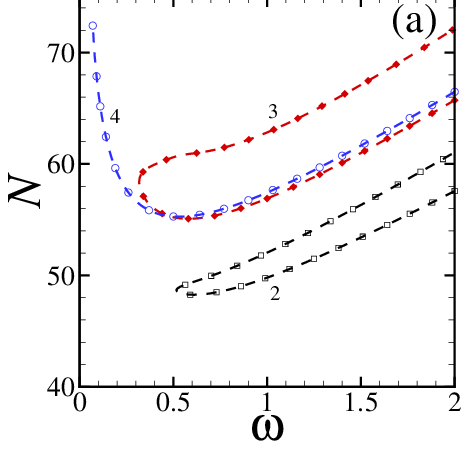,width=5cm}%
\epsfig{file=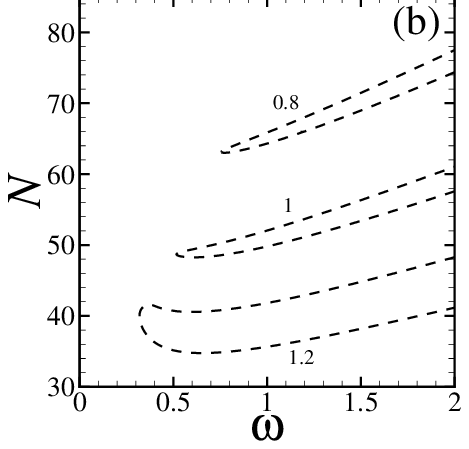,width=5cm}
\caption{(Color online) The same as in Fig. \protect\ref{vort_exist}, but
for asymmetric vortex solutions. The colors and symbols in (a) allow to
follow the overlapping existence curves.}
\label{vort_exist_as}
\end{figure}

According to the linear stability analysis (also verified by the direct
simulations), symmetric and asymmetric vortices with different $l$ and $%
\Delta W$ are unstable. This may be explained by the fact that the spread of
the nonlinearity is not sufficient for the creation of stable discrete
vortices, whose stability region is narrow in the model with the uniform
onsite nonlinearity \cite{vortex}.

\section{Conclusion}

In this work we have introduced the 2D lattice model, based on the linear
discrete Schr\"{o}dinger equation with two nonlinear sites embedded into it,
and a generalization for the moderately localized nonlinearity with two
separated maxima. The setting may be implemented in arrays of optical
waveguides, and in BEC trapped in a deep optical-lattice potential. The
existence and stability of discrete 2D solitons of the symmetric,
antisymmetric, and asymmetric types have been considered in these systems.
The setting with the moderate localization of the nonlinearity admits a much
greater variety of the localized modes. The subcritical bifurcation, which
accounts for the transition from symmetric to asymmetric modes, has been
identified. In the case of the moderate localization, the existence and
stability of discrete symmetric and asymmetric vortices have been also
discussed.


\begin{thebibliography}{99}
\bibitem{book} P. G. Kevrekidis, \textit{The Discrete Nonlinear Schr\"{o}%
dinger Equation: Mathematical Analysis, Numerical Computations, and Physical
Perspectives} (Springer: Berlin and Heidelberg, 2009).

\bibitem{Tsironis} M. I. Molina, G. P. Tsironis, Phys. Rev. B \textbf{47}
(1993) 15330 (1993); B. C. Gupta, K. Kundu, Phys. Rev. B \textbf{55} (1997)
894.

\bibitem{we} V. A. Brazhnyi, B. A. Malomed, Phys. Rev. A \textbf{83} (2011)
053844.

\bibitem{Almas} B. Maes, M. Solja\v{c}i\'{c}, J. D. Joannopoulos, P.
Bienstman, R. Baets, S.-P. Gorza, M. Haelterman, Opt. Express, \textbf{14}
(2006) 10678; E. N. Bulgakov, A. F. Sadreev, Phys. Rev. B \textbf{81} (2010)
115128; E. Bulgakov, A. Sadreev, K. N. Pichugin, \textit{ibid}. B \textbf{83}
(2011) 045109.



\bibitem{big-review} F. Lederer, G. I. Stegeman, D. N. Christodoulides, G.
Assanto, M. Segev, Y. Silberberg, Phys. Rep. \textbf{463} (2008) 1.

\bibitem{plasmonics} A. A. Sukhorukov, A. S. Solntsev, S. S. Kruk, D. N.
Neshev, and Y. S. Kivshar, Opt. Lett. \textbf{39} (2014) 462.

\bibitem{BEC} A. Trombettoni, A. Smerzi, Phys. Rev. Lett. \textbf{86}, 2353
(2001); A. Smerzi, A. Trombettoni, Phys. Rev. A \textbf{68}, 023613 (2003);
F. Kh. Abdullaev, B. B. Baizakov, S. A. Darmanyan, V. V. Konotop, M.
Salerno, Phys. Rev. A \textbf{64}, 043606 (2001); A. Smerzi, A. Trombettoni,
P. G. Kevrekidis, A. R. Bishop, Phys. Rev. Lett. \textbf{89}, 170402 (2002);
V. Ahufinger1 A. Sanpera, P. Pedri, L. Santos, M. Lewenstein, Phys. Rev. A
\textbf{69} (2004) 053604; V. A. Brazhnyi, V. V. Konotop, V. M. P\'{e}%
rez-Garc\'{\i}a, \textit{ibid}. \textbf{74} (2006) 023614.

\bibitem{BEC-review} D. Jaksch, P. Zoller, Ann. Phys. \textbf{315} (2005)
52; M. Lewenstein, A. Sanpera, V. Ahufinger, B. Damski, A. Sen, U. Sen,
Advances in Physics \textbf{56} (2007) 243; B. Ilan, M. I. Weinstein,
Multiscale Model. Simul. \textbf{8} (2010) 1055.

\bibitem{Engineering-OL} P. Windpassinger, K. Sengstock, Rep. Prog. Phys.
\textbf{76} (2013) 086401.

\bibitem{Edwin} B. A. Malomed, E. Ding, K. W. Chow, S. K. Lai, Phys. Rev. E
\textbf{86}, 036608 (2012); K. W. Chow, E. Ding, B. A. Malomed, A. Y. S.
Tang, Eur. Phys. J. Special Topics \textbf{223 }(2014) 63.

\bibitem{Feshbach} S. Giorgini, L. P. Pitaevskii, and S. Stringari, Rev.
Mod. Phys. \textbf{80} (2008) 1215; H. T. C. Stoof, K. B. Gubbels, D.
B. M. Dickrsheid, \textit{Ultracold Quantum Fields} (Springer: Dordrecht,
2009).

\bibitem{tiltedOLandFeshbach} 
E. D\'{\i}az, C. Gaul, R. P. A. Lima, F. Dom%
\'{\i}nguez-Adame, and C. A. M\"{u}ller, Phys. Rev. A \textbf{81} (2010)
051607(R); G. A. Sekh, Phys. Lett. A \textbf{376} (2012) 1740.

\bibitem{Yb} R. Yamazaki, S. Taie, S. Sugawa, Y. Takahashi, Phys. Rev. Lett.
\textbf{105} (2010) 050405.

\bibitem{Herring} G. Herring, P. G. Kevrekidis, B. A. Malomed, R.
Carretero-Gonz\'{a}lez, D. J. Frantzeskakis, Phys. Rev. E \textbf{76} (2007)
066606.

\bibitem{Ljupco} Lj. Had\v{z}ievski, G. Gligori\'{c}, A. Maluckov, and B. A.
Malomed, Phys. Rev. A \textbf{82} (2010) 033806.

\bibitem{bif} G. Iooss and D. D. Joseph, \textit{Elementary Stability
Bifurcation Theory} (Springer-Verlag: New York, 1980).

\bibitem{we2} V. A. Brazhnyi, B. A. Malomed,
%\textit{Symmetric and asymmetric localized modes in linear lattices with
%an embedded pair of $\chi ^{(2)}$nonlinear sites},
Phys. Rev. A \textbf{86} (2012) 013829.

\bibitem{Thawatchai} T. Mayteevarunyoo, B. A. Malomed, G. Dong, Phys. Rev. A
\textbf{78} (2008) 053601 (2008); N. Dror, B. A. Malomed, \textit{ibid.} A
\textbf{83} (2011) 033828 (2011); X.-F. Zhou, S.-L. Zhang, Z.-W. Zhou, B. A.
Malomed, H. Pu, \textit{ibid}. A \textbf{85} (2012) 023603.

\bibitem{Ady} A. Shapira, N. Voloch-Bloch, B. A. Malomed, A. Arie, J. Opt.
Soc. Am. B \textbf{28} (2011) 1481.

\bibitem{JModOpt} T. Mayteevarunyoo, B. A. Malomed, A. Reoksabutr, J. Mod.
Opt. \textbf{58} (2011) 1977.

\bibitem{Barcelona} Y. V. Kartashov, B. A. Malomed, Torner, Rev. Mod. Phys.
\textbf{83} (2011) 247.

\bibitem{Jena} A. Szameit, J. Burghoff, T. Pertsch, S. Nolte, A. T\"{u}%
nnermann, Lederer, Opt. Exp. \textbf{14} (2006) 6055; A. Szameit, S. Nolte,
J. Phys. B: At. Mol. Opt. Phys. \textbf{43} (2010) 163001.

\bibitem{doping} J. Hukriede, D. Runde, D. Kip, J. Phys. D: Appl. Phys.
\textbf{36} (2003) R1.

\bibitem{Yang2008} Z. Shi, J. Wang, Z. Chen, and J. Yang, Phys. Rev. A \textbf{78} (2008) 063812.

\bibitem{Dohnal2009} T. Dohnala, H. Uecker, Physica D \textbf{238} (2009) 860.


\bibitem{VK} M. Vakhitov, A. Kolokolov, Radiophys. Quantum. Electron.
\textbf{16} (1973)783 (1973); L. Berg\'{e}, Phys. Rep. \textbf{303} (1998)
259.

\bibitem{vortex} B. A. Malomed and P. G. Kevrekidis, Phys. Rev. E \textbf{64} (2001) 026601.



\end{thebibliography}
\end{document}